\documentstyle[preprint,aps]{revtex}

\begin{document}

\draft

\title{On the integrability of halo dipoles in gravity}

\author{Werner M. Vieira\thanks{e-mail: vieira@ime.unicamp.br}
and
 Patricio S. Letelier\thanks{e-mail: letelier@ime.unicamp.br}}
\address{Departamento de Matem\'atica Aplicada\\
Instituto de Matem\'atica, Estat\'{\i}stica e Ci\^encias da
 Computa\c{c}\~ao\\
Universidade Estadual de Campinas, CP 6065\\
13081-970 Campinas, SP, Brazil\\
{\rm September 17, 1996 }}

\maketitle

\begin{abstract}

We stress that halo dipole components are nontrivial in
core--halo systems in both
Newton's gravity and General Relativity.
To this end, 
we extend a recent exact relativistic model
to include also a halo dipole component. Next,
we consider orbits evolving in the inner vacuum between a
monopolar core and a pure halo dipole and find that, while the
Newtonian dynamics is integrable, its relativistic counterpart
is chaotic. This shows
that chaoticity due only to
halo dipoles is an intrinsic relativistic gravitational effect.

\end{abstract}


We report in this Letter a simple situation
where the contact between Newton's gravity and
General Relativity (GR)
allows for a clear distinction between their
predictions about chaotic behavior of orbit dynamics.

We considered in \cite{WernerPatricio} how does GR
deal with a monopolar core (at first approximation),
which may be a neutron star, a galactic nucleus 
or even a black hole, surrounded by a distant, massive halo
of dust. We exhibited an exact, static, axially symmetric
model with a halo made of quadrupoles and
octopoles of arbitrary strengths. In constructing this
solution, we were also guided
by its formal resemblance to the classical
H\'enon--Heiles potential in the Newtonian limit.

Here, we are mainly interested in the
dynamical rather than formal
comparison between Newtonian core--halo systems
and its relativistic counterpart.
Firstly, we state the problem
according to Newton's theory: let the coordinate origin stay
at the mass center of the monopolar core,
$z$ be the symmetry axis
of the core--halo system and $D$ be the region between the
minimum and the maximum spheres centered at the origin
that isolate the inner vacuum from the 
core and the halo. We have to solve
Laplace's  equation in $D$ for the axially
symmetric Newtonian potential.
By using standard spherical multipolar expansion,
we arrive at the following gravitational potential
felt by test particles evolving in $D$:
\begin{equation}
\Phi_N=\Phi_0-\frac{1}{\rho}-{\cal{D}}z
+\frac{{\cal{Q}}}{2}(2z^2-r^2)
+\frac{{\cal{O}}}{2}(2z^3-3zr^2) + \cdots
{\mbox{\hspace{0.4cm}.}}\label{1}
\end{equation}
$\Phi_0$ is a constant 
(put here only for the sake of completeness)
and $\rho^2=r^2+z^2=x^2+y^2+z^2$
where $x,y,z$ are the usual Cartesian coordinates.
$\Phi_N$ and all quantities in it are written in
convenient nondimensional units.
${\cal{D}}$, ${\cal{Q}}$ and ${\cal{O}}$ are respectively 
the dipole, quadrupole and octopole halo strengths given by
the appropriate normalized integrals over the halo
(as a source) in the Newtonian approach. It is worth to
stress that, quite apart from the nomenclature,
the halo multipoles have opposite
behavior in $D$ from the usual decreasing--with--distance core
multipoles.
In particular, by merely
translating the origin to the core mass center
we can (and in fact we do) eliminate
the usual core dipole in $D$, so that the next
higher core contribution should be
the usual quadrupolar one (which will not be considered here).
On the other hand, nothing so simple can eliminate
the halo dipole component. As it is easily seen, the
permanency of the later is related to the
halo mass distribution
with respect to the plane $z=0$. The only way of
eliminating halo dipoles from (\ref{1})
should be passing to a noninertial reference
frame with uniform acceleration ${\cal{D}}$ in
the $z$--direction. In this case,
we should have to properly treat
this problem as a relativistic one,
since it extrapolates the Galilean
gauge invariance of the Newtonian physics.

Now, we consider bounded orbits under the Newtonian potential
(\ref{1}). If we switch the halo quadrupole and octopole off
(${\cal{Q}}={\cal{O}}=0$), thus the remaining potential
describing the core plus the pure halo dipole
is integrable, in particular being separable in parabolic
coordinates~\cite{GDRH}. The hamiltonian $H$ as well as the
additional constant of motion $C$
in involution with the hamiltonian are
\begin{eqnarray}
H=\frac{1}{2}(p_r^2+p_z^2)+\frac{\ell^2}{2r^2}+\Phi_N,\nonumber\\
C=rp_rp_z-zp_r^2-\ell^2\frac{z}{r^2}+\frac{z}{\rho}-
\frac{{\cal{D}}}{2}r^2,
\label{3}
\end{eqnarray}
where $\ell$ is the nondimensional conserved
angular momentum associated to the axial symmetry.
Only to illustrate the integrable dynamics,
we present in Fig.\ 1(a) a typical $z=0$
Poincar\'e section for this case.
Contrasting with this, we mention that either the
Newtonian halo quadrupole (${\cal{Q}}\neq 0$)
or octopole (${\cal{O}}\neq 0$)
does produce chaos, which will be shown in
an extended forthcoming paper.

Since the Newtonian halo dipole is nontrivial in $D$,
appearing quite naturally in (\ref{1}),
we ask for its relativistic counterpart. In fact, we succeeded
in extending our previous relativistic
solution~\cite{WernerPatricio}
to add a dipole term to the halo. Starting from Weyl's metric
and following the formulation presented therein,
the solution depends upon
two functions $\nu(u,v)$ and $\gamma(u,v)$, the last being
obtained by quadrature from the former one.
Here, it is sufficient
to write down only the main function 
$\nu(u,v)$ (see~\cite{WernerPatricio} for notation and details):
\begin{equation}
\nu(u,v)=a_0Q_0(u)-{\cal{D}}P_1(u)P_1(v)
+(2/3){\cal{Q}}P_2(u)P_2(v)+(2/5){\cal{O}}P_3(u)P_3(v).
\label{2}
\end{equation}
This equation should be compared with eq.\ (3) of
that Reference.
Essentially, the 
first term with  $a_0=-1$ describes the monopolar core
(that can be switched off by putting $a_0=0$),
the remaining ones being the corresponding
multipoles originated from the halo. $Q_0$ and $P_n$ are the
corresponding Legendre's functions in standard notation
and $u,v$ are prolate spheroidal coordinates constructed
from Weyl's ones.

What is the Newtonian limit of (\ref{2})?
The basic steps to obtain it are:
i) The relativistic solution is 
valid in the full vacuum between the core and the 
halo. ii) We assume that there exist a region
$\overline{D}$ in the 
intermediate vacuum where the conditions of weak gravitational 
field and slow motion of test particles occur. Then, Eintein's 
equations reduce in $\overline{D}$ to Laplace's
equation for the Newtonian 
potential $\Phi_N$, which relates to the 
metric through the temporal component $g_{tt}=1+(2/c^2)
\Phi_N$ (the remaining spatial components
of the metric being irrelevant to this approximation).
iii) The Schwarzschild coordinates at which we finally
come back with the solution approximate to the usual
Euclidean spherical 
ones plus time in $\overline{D}$. We consistently
assume that 
this region is far from the core's event horizon. Now, by 
applying all this to the relativistic solution (\ref{2})
and expanding $g_{tt}$ to the first order in
${\cal{D}}$, ${\cal{Q}}$ and ${\cal{O}}$
we arrive, after the conversion to
nondimensional units, exactly
at the potential $\Phi_N$ given by (\ref{1})
(apart from the irrelevant constant $\Phi_0$).
The implicit assumption here is that
$\overline{D}$ and the Newtonian region $D$ above
have a nonempty intersection at least.
We would like to stress
the nontrivial character of halo dipoles
also in the relativistic context, as they contribute
to the Kretschmann's scalar
$R^{\alpha \beta \gamma \delta}R_{\alpha \beta \gamma \delta}$
even if $a_0={\cal{Q}}={\cal{O}}=0$.
Moreover, they are
smooth, however exact, relativistic counterparts
of Newtonian halo dipoles, in the sense that
their first contribution
to the Riemann tensor is of second order in ${\cal{D}}$.

Next, we compare bounded geodesics generated by (\ref{2})
for a pure halo dipole
(${\cal{Q}}={\cal{O}}=0$) with the preceding Newtonian case.
A typical Poincar\'e section of relativistic orbits
is shown in Fig.\ 1(b) where vast chaotic layers are
seen in the geodesic motion. This should be
contrasted with the full integrability of
the corresponding Newtonian figure 1(a).
Since Poincar\'e sections are global, gauge
invariant tools to detect chaos in bounded systems,
this proves that chaoticity in the vacuum between a monopolar
core and a purely dipolar
halo of dust is an intrinsic relativistic gravitational effect.

An extended numerical study of
both newtonian and relativistic cases
as well as the details of the
full relativistic solution (\ref{2}) will
be presented in an extended forthcoming paper.

The authors thank CNPq and FAPESP for financial support.

\begin{figure}

\caption{Poincar\'e sections of orbits
in the vacuum between a monopolar core and a pure halo dipole (${\cal{Q}}={\cal{O}}=0$) through the plane $z=0$ (for $p_z>0$):
(a) the Newtonian case in Euclidean cilyndrical coordinates
for the potential (\protect\ref{1}) with
nondimensional orbit parameters
$E=-0.03$ (energy), $\ell =3.8$ (angular momentum) and
${\cal{D}}=5.\times 10^{-4}$ (halo dipole strength), and
(b) the relativistic case for timelike geodesic motion
in Weyl's coordinates for the
solution  (\protect\ref{2}) with $E=0.975$, $\ell =3.8$ and
${\cal{D}}=3.\times 10^{-4}$. Note that the
relativistic energy
$E=0.975$ corresponds in the Newtonian limit to the energy
$E=-0.025$.}

\end{figure}


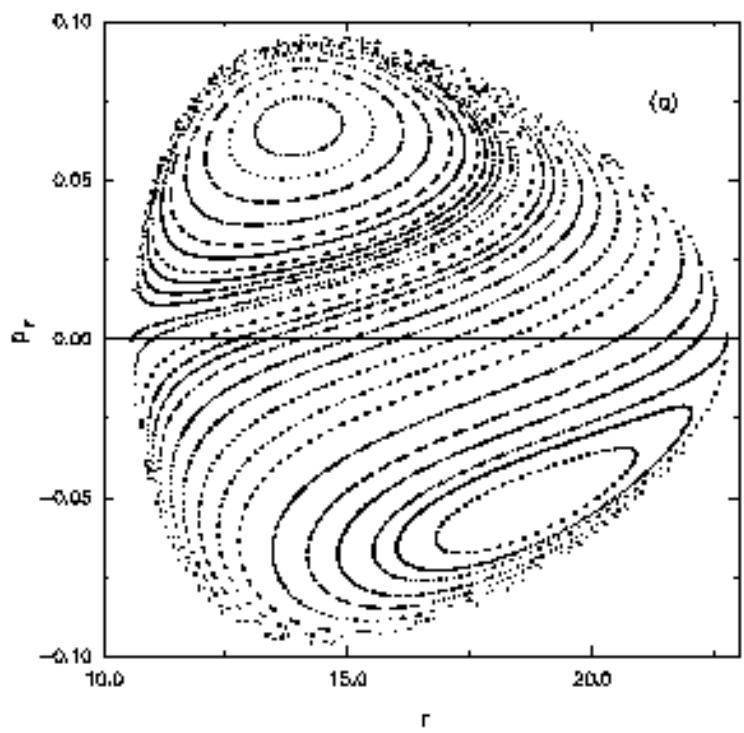
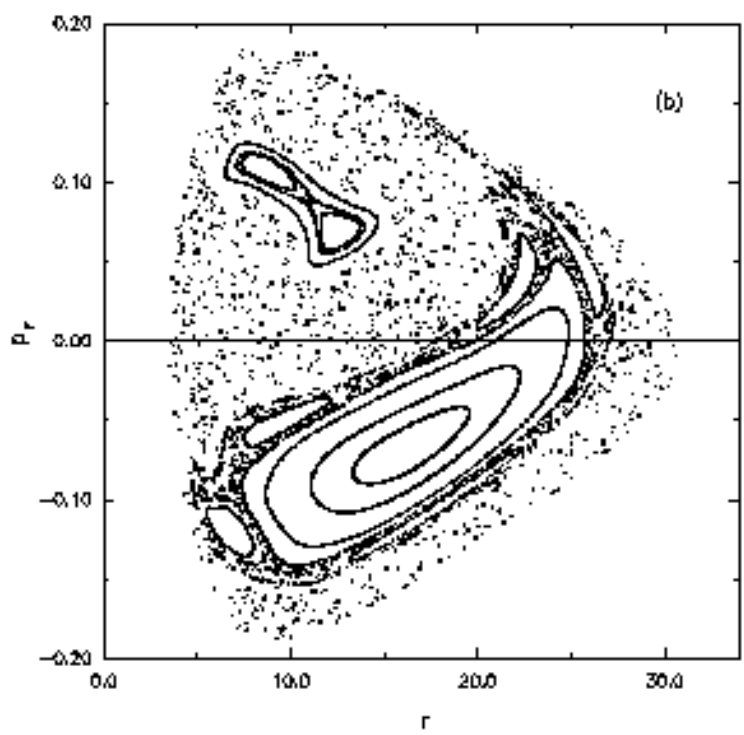
\end{document}